\documentclass[aps,twocolumn,floatfix,showpacs]{revtex4}
\usepackage{epsfig}
\newcommand{\be}{\begin{equation}}
\newcommand{\ee}{\end{equation}}

\begin{document}

\title{Data Compression and Entropy Estimates by Non-sequential Recursive Pair Substitution}

\author{Peter Grassberger}

\affiliation{John-von-Neumann Institute for Computing, Forschungszentrum J\"ulich,
D-52425 J\"ulich, Germany}

\date{\today}

\begin{abstract}
We argue that Non-sequential Recursive Pair Substitution (NSRPS) as suggested by 
Jim\'enez-Monta\~no and Ebeling can indeed be used as a basis for an optimal data 
compression algorithm. In particular, we prove for Markov sequences that NSRPS together 
with suitable codings of the substitutions and of the substitute series does not lead 
to a code length increase, in the limit of infinite sequence length. When applied 
to written English, NSRPS gives entropy estimates which are very close to those obtained 
by other methods. Using ca. 135 GB of input data from the project Gutenberg, we estimate
the effective entropy to be $\approx 1.82$ bit/character. Extrapolating to infinitely 
long input, the true value of the entropy is estimated as $\approx 0.8$ bit/character.
\end{abstract}


\maketitle

\section{Introduction}

The discovery that the amount of information in a message (or in any other structure)
can be objectively measured was certainly one of the major scientific achievements of 
the 20th century. On the theoretical side, this quantity -- the information theoretic 
entropy -- is of interest mainly because of its close relationship to thermodynamic
entropy, its importance for chaotic systems, and its role in Bayesian 
inference (maximum entropy principle). Practically, estimating 
the entropy of a message (text document, picture, piece of music, etc.) is important 
because it measures its compressibility, i.e. the optimal achievement for any 
possible compression algorithm. In the following, we shall always deal with sequences
$(s_0,s_1,\ldots)$ built from the characters of a finite alphabet $A = \{a_0,\ldots,a_{m-1}\}$
of size $m$. In the simplest case the alphabet consists just of 2 characters, in 
which case the maximum entropy is 1 bit per character.

Indeed, information entropy as introduced by Shannon \cite{shannon} is a probabilistic 
concept. It requires a measure (probability distribution) to be defined on the set 
of all possible sequences. In particular, the probability for $s_t$
to be given by $a_k$, given all characters $s_0,s_1,\ldots, s_{t-1}$, is given by
\begin{eqnarray}
   p_t(k|k',k'',\ldots) = &&\\{\rm prob}(s_t = a_k&|&s_{t-1}=a_{k'}, s_{t-2}=a_{k''}, \ldots
     ) . \nonumber
\end{eqnarray}
In case of a stationary measure with finite range correlations, $p_t(k|k',k'',\ldots)$ 
becomes independent of $t$ for $t\to\infty$. Then Shannon's famous formula,
\be
   h = \lim_{i\to \infty} h^{(i)}
\ee
with 
\be
   h^{(i)} = - \sum_{k_1\ldots k_i} p(k_1\ldots k_i) \log_2 p(k_1|k_2\ldots k_i)\;,
\ee
gives the {\it average} information per character. The generalization to non-stationary 
measures is straightforward but will not be discussed here.

In contrast to this approach are attempts to define the {\it exact} information content of a single 
finite sequence. Theoretically, the basic concept here is the {\it algorithmic complexity} 
AC (or algorithmic {\it randomness}) \cite{kolmogorov,chaitin}. For any given universal 
computer $U$, the AC of a sequence $S$ relative to $U$ is given by
the length of the shortest program which, when input to $U$, prints $S$ and then makes 
$U$ to stop, so that the next sequence can be read. If $S$ is randomly drawn from 
a stationary ensemble with entropy $h$, then one can show that the AC
per character tends towards $h$, for almost all $S$ and all $U$, as the length of $S$
tends towards infinity \cite{li-vitanyi}. Thus, except for rare sequences which do not 
contribute to averages, $h$ sets the limit for the compressibility.

Practically, the usefulness of AC is limited by the fact that there cannot exist any 
algorithm which finds for each $S$ its shortest code (such an algorithm could be used to 
solve Turing's halting problem, which is known to be impossible) \cite{li-vitanyi}. 
But one can give algorithms which are often quite efficient. Huffman,
arithmetic, and Lempel-Ziv coding are just three well known examples \cite{cover}.
Any such algorithm can be used to give an upper bound to $h$ (modulo fluctuations from 
the finite sequence length) while, inversely, knowledge of $h$ sets a lower limit to 
the average code lengths possible with these codes.

A data compression scheme is called {\it optimal}, if it does not do much worse than the 
best possible for typical random strings. More precisely, let $\{S\}$ be a set of sequences
with entropy $h(S)$, and let the code string $C(S)$ be built from an alphabet of $m_C$
characters. Then we call the coding scheme $C: S\to C(S)$ optimal, if 
\be
   {{\rm length}[C(S)] \over {\rm length}[S]} \to {h \over \log_2 m_C }
                            \quad {\rm for} \;\; {\rm length}[S] \to \infty
\ee
and for nearly all $S$.
While Huffman coding is not optimal, arithmetic and Lempel-Ziv codings are \cite{cover}.

In several papers, Jim\'enez-Monta\~no, Ebeling, and others \cite{jimenes,poeschl} have 
suggested coding schemes by non-sequential recursive pair substitution (NSRPS) \cite{footnote0}. 
Call the original sequence $S_0$. We count the numbers $n_{jk}$ of non-overlapping successive 
pairs of characters in $S_0$ where $s_t = a_j$ and $s_{t+1} = a_k$, and find their maximum,
$n_{\rm max} = \max_{j,k< m} n_{jk}$. The corresponding index pair is $(j_0,k_0)$. 
Then we introduce a new character by concatenation
\be
   a_m = (a_{j_0}a_{k_0})
\ee
and form the sequence $S_1$ by replacing everywhere the pair $a_{j_0}a_{k_0}$ by $a_m$. For 
the special case of $j_0 = k_0$, any string of $2r+1$ characters $a_{j_0}$ is replaced by $r$ 
characters $a_m$, followed by one $a_{j_0}$.

This is then repeated recursively: The sequence $S_{i+1}$ is obtained from $S_i$ by replacing 
the most frequent pair $a_{j_i}a_{k_i}$ by a new character $a_{m+i}$. The procedure stops
if one can argue that further replacements would not possibly be of any use. Typically this 
will happen if the code length consisting of both a description of $S_{i+1}$ and a description
of the pair $(j_i,k_i)$ is definitely longer than a description of $S_i$, for the present 
and all subsequent $i$.

Thus one sees that efficient encodings (which must also be uniquely decodable!) of the 
sequences $S_i$ and of the type of substituted pairs become crucial for the analysis of NSRPS. 
Unfortunately, the ``codings" given in \cite{jimenes,poeschl} are neither efficient nor 
uniquely decodable \cite{footnote}. Thus their ``complexities" have no direct relationship 
to $h$ or to algorithmic complexity (in contrast to their claim), and it is not clear from 
their work whether NSRPS can be made into an optimal coding scheme at all.

It is the purpose of the present paper to give at least partial answers to this.
More precisely, we shall only be concerned with the limit of infinitely long 
strings, where the information encoded in the pairs $(j_i,k_i)$ can be neglected
in comparison with the information stored in $S_i$, at least for any finite $i$.
We will first show analytically that a coding scheme for $S_i$ exists which 
satisfies a necessary condition for optimality (Sec.2). We then apply this to written 
English (Sec.3), where we shall also compare our estimates of $h$ to those obtained 
with other methods.

\section{NSRPS for Markov sequences}

Let us for the moment assume that $S_0$ is binary (the two characters are ``0" and ``1"),
and that it is completely random, i.e. identically and independently distributed (iid) 
with the same probability for each character. Thus $p(0|\ldots) = p(1|\ldots) = 1/2$, and 
$h=1$ bit. The length of $S_0$ is $N_0$, thus the total average information 
stored in $S_0$ is $N_0$ bits.

No coding scheme can reduce the length of $C(S_0)$ to less than $N_0$ bits
on average. Indeed, all schemes will have ${\rm length}[C(S_0)] > N_0$ bits (strict
inequality!), unless the ``coding" is a verbatim copy. For a coding scheme to be 
optimal, a necessary (but not sufficient) condition is that 
\be 
   {\rm length}[C(S_0)] / N_0 \to 1 \;{\rm bit}
\ee
for $N_0\to\infty$, i.e. the overhead in the code must be less than extensive
in the sequence length. This is what we want to show here, together with its 
generalization to arbitrary (first order) Markov sequences.

For this, we need two lemmata:

{\bf Lemma 1}: {\it For any Markov sequence $S_0$ (not necessarily binary, and not 
necessarily iid) built from $m$ letters, the sequence $S_1$ is again Markov.}

{\bf Lemma 2}: {\it If a word $w = (k,k',k'',\ldots)$ appears several times in $S_0$,
and if one of these instances is substituted in $S_i$ by a string of characters
not straddling its boundaries, then all other instances of $w$ in $S_0$ are also
substituted in $S_i$ by the same string.}

Lemma 1 tells us that NSRPS might make the structure of $S_i$ more complex than 
that of $S_0$, but not much so. Being a Markov chain, its entropy can be estimated
if the transition probabilities $p(k|k_1)$ are known. Thus estimating the entropy 
of $S_i$ reduces to estimating di-block entropies $h^{(2)}$, which is straightforward (at 
least in the limit $N_0\to\infty$).

Lemma 2 tells us that there cannot be any ambiguity in $S_i$. In particular, 
it cannot happen that more information is needed to specify $S_i$ than there 
is needed to specify $S_0$, since the mapping $S_0\to S_i$ is bijective, once 
the substitution rules are fixed.

The proofs of the lemmata are easy. Let us denote by $p_j(\ldots)$ the probability 
distributions after $j$ pair substitutions. For lemma 1 we just have to show that 
$p_1(k|k',k'')$ is independent of $k''$ for each pair $(k,k')$, provided the 
same holds also for $p_0$. This follows basically from the fact that any 
substitution makes the sequence shorter. But the detailed proof is somewhat 
tedious, because $p_1(k|k',k'')\neq p_0(k|k',k'')$, even if all $k$'s are less than 
$m$, $k\neq k_0$, $k''\neq j_0$, and neither $(k,k')$ nor $(k',k'')$ are equal to 
the pair $(j_0,k_0)$. In that case,
$(N_0-n_{\rm max}) p_1(k|k',k'') = N_0 p_0(k|k',k'')$, and independence of $k''$ follows
immediately. All other cases have to be dealt with similarly. For instance, if 
either $(k,k')$ or $(k',k'')$ is the pair $(j_0,k_0)$, then $p_1(k,k',k'')=0$. Else,
if $k''=m\neq k,k'$, then $p_1(k|k',k'') = N_0/ (N_0-n_{\rm max}) p_0(k|k',j_0,k_0) = 
N_0/ (N_0-n_{\rm max}) p_0(k|k')$. We leave the other cases as exercises to the reader.

For proving lemma 2 we proceed indirectly. We assume that there is a word in 
$S_0$ which is encoded differently in different locations. Let us assume 
that this difference happened for the first time after $i$ substitutions.
Since only one type of pair is exchanged in each step, this means that a 
substitution is skipped in one of the locations, at this step. But this is 
impossible, since {\it all} possible substitutions are made at each step.

From the two lemmata we obtain immediately our central 

{\bf Theorem:} {\it If $S_0$ is drawn from a (first order) Markov process with length $N_0$ 
and entropy $h_0 =  - \sum_{k,k'} p_0(k,k') \log_2 p_0(k|k')$, then every $S_i$ 
is also Markovian in the limit $N_0\to\infty$, with entropy 
\be
   h_i = h_i^{(2)} =  - \sum_{k,k'} p_i(k,k') \log_2 p_i(k|k')                \label{h2}
\ee
and with length $N_i$ satisfying $N_i/N_0 = h_0/h_i$.}

Thus the total amount of information needed to specify $S_i$ is the same as that for 
$S_0$, for infinitely long sequences. Since the overhead needed to specify the pairs 
$(j_i,k_i)$ can be neglected in this limit, we see that we do not loose code length 
efficiency by pair substitution, provided we take pair probabilities correctly 
into account during the coding. The actual encoding can be done by means of an arithmetic
code based on the probabilities $p_i(k|k')$ \cite{cover}, but we shall not work out 
the details. It is enough to know that the code length then becomes equal to the 
information (both measured in bits), for $N_0\to\infty$.

Let us see in detail how all this works for completely random iid binary 
sequences. The original sequence $S_0 = 00101001111010011011\ldots$ 
has $p_0(00)=p_0(01)=p_0(10)=p_0(11)=1/4$ and therefore $h_0 = 1$ bit. Thus 
we can, without loss of generality, assume that the new character is 
$2 = (01)$, so that $S_1 = 02202111202121\ldots$. The 3 characters are 
now equiprobable, $p_1(0)=p_1(1)=p_1(2)=1/3$, but they are not independent
since of course $p_1(01)=0$. Indeed, one finds $p_1(00)=p_1(02)=p_1(11)=p_1(21)=1/6,
\;p_1(10)=p_1(12)=p_1(20)=p_1(22)=1/12$. The order-2 entropy of $S_1$ is easily
calculated as $h_1^{(2)} = 4/3 \log_2 2$. On the other hand, since $N_0/4$ pairs 
have been replaced by single characters, the length of $S_1$ is $N_1=3N_0/4$. Thus,
if $S_1$ is Markov, then the total information needed to specify it is 
$N_1 h_1^{(2)} = N_0$ bits, the same as for $S_0$. If it were not Markov, its
information would be smaller. But this cannot be, because the map $S_0\to S_1$ 
was invertible. Thus $S_1$ must indeed be Markov, as can also be checked explicitly.

In the next step, we can either replace $(21) \to 3$ or $(02) \to 3$, since both 
have the same probability. If we do the former, the sequence becomes 
$S_2 = 02203112033\ldots$. Now the letters are no longer equiprobable,
$p_2(1)=p_2(2)=p_2(3)=1/5$, $p_2(0)=2/5$. Calculating $N_2, p_2(kk')$, and 
$h_2^{(2)}$ is straightforward, and one finds again $N_2 h_2^{(2)} = N_0$ bits. 
Thus one concludes that $S_2$ must also be Markov. For the next 
few steps one can still verify 
\be
   N_i h_i^{(2)} = \ldots N_0 \; {\rm bits},           \label{same}
\ee
by hand, but this becomes increasingly tedious as $i$ increases.

\begin{figure}
%Fig 1
\psfig{file=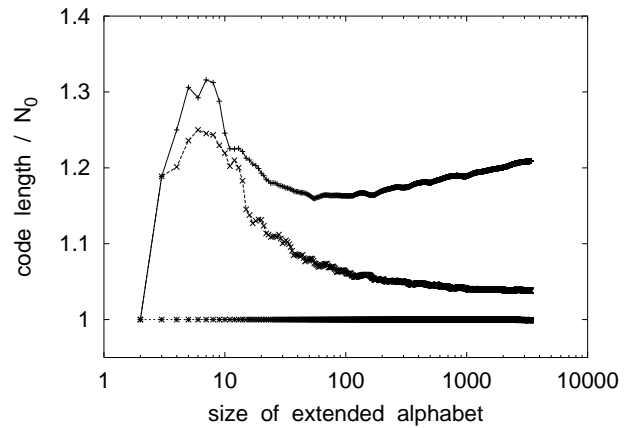,width=5.8cm,angle=270}
\caption{Results for a completely random (iid, uniformly distributed) binary
  initial sequence of $N_0 = 8\times 10^8$ bits, plotted against the size of 
  the extended alphabet. Uppermost curve: code length needed to encode $S_i$, 
  divided by $N_0$, if $\log_2 (i+2)$ bits are used for each character. Middle 
  curve: code length based on $h_i^{(1)}$, i.e. the single-character distributions 
  $p_i(k)$ are used in the encoding. Lowest curve, indistinguishable on this
  scale from a horizontal straight line: 
  code length based on $h_i^{(2)}$, using the two-character distributions $p_i(k,k')$.}
\label{fig1.ps}
\end{figure}

Thus we have verified Eq.(\ref{same}) by extensive simulations, where we found 
that it is exact, within the expected fluctuations, up to several thousand 
substitutions (Fig.1). The distribution of the probabilities $p_i(k)$ becomes very 
wide for large $i$, i.e. the sequences $S_i$ are far from uniform for large $i$, 
but they are Markov and their entropies $h_i^{(2)}$ are exactly (within 
the expected systematic finite sample corrections \cite{herzel,grass-fsc})
equal to $N_0/N_i$ bits. Notice that if we would encode the last $S_i$ without
taking the correlations into account (as seems suggested in
\cite{jimenes,poeschl}), then the code length for it would be larger and the 
coding scheme would not be optimal.

We have also made some simulations where we started with non-trivial Markov 
processes for $S_0$, or even with non-Markov sequences with known entropy. 
The latter were generated by creating initially a binary iid sequence with 
$p(0) \neq p(1)$, and then using this as an input configuration for a few 
iterations of the bijective cellular automaton R150 (in Wolfram's notation)
\cite{sg}.

\begin{figure}
%Fig 2
\psfig{file=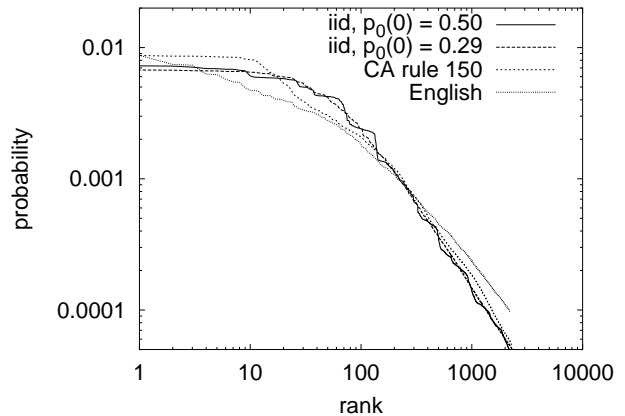,width=5.8cm,angle=270}
\caption{Ranked single character probability distributions $p_i(k)$ of strings after 
  $i=2298$ pair substitutions. The different curves are for a completely random iid
  initial string $S_0$ (solid line), iid string $S_0$ with $p_0(0)=0.29$ (long dashed), 
  $S_0$ obtained by applying two times CA rule 150 to an iid sequence with 
  $p(0)=0.09$ (dashed), and to written English with a reduced (46 character) 
  alphabet (dotted).}
\label{fig2.ps}
\end{figure}

From these simulations it seems that $N_i h_i^{(2)}$ always tends towards $N_0$.
Also, the probability distributions $p_i(k)$ seem to tend (very slowly, see 
Fig.2) to the same scaling limit as for iid and uniform $S_0$. This suggests
that indeed $S_i$ tends to a Markov process for arbitrary $S_0$. In this
case an optimal coding would be obtained if one would use, e.g., an 
arithmetic code to encode $S_i$ by using approximate values of the observed 
$p_i(k|k')$ for large $i$.

Thus we have given strong (but still incomplete) arguments that NSRPS combined 
with efficient coding of $S_i$ gives indeed an optimal coding scheme. In 
practice, it would of course be extremely inefficient in terms of speed, and 
thus of no practical relevance. But it could well be that it might lead to 
more stringent entropy estimates than other methods. To test this we shall
now turn to one of the most complex and interesting system, written natural
language.

\section{The entropy of written English}

The data used for the application of NSRPS to entropy estimation of written 
English consisted of ca. 150 MB of text taken from the Project Gutenberg 
homepage \cite{gutenberg}. It includes mainly English and American novels
from the 19th and early 20th century (Austen, Dickens, Galsworthy, Melville, 
Stevenson, etc.), but also some technical reports (e.g. Darwin, historical 
and sociological texts, etc.), Shakespeares collected works, the King James 
Bible, and some novels translated from French and Russian (Verne, Tolstoy,
Dostoevsky, etc.). 

From these texts we removed first editorial and legal remarks added by the 
editors of Project Gutenberg. We also removed end-of-line, end-of-page, and 
carriage return characters. All runs of consecutive blanks were replaced by 
a single blank. Finally, we also removed all characters not in 
the 7-bit ASCII alphabet (ca. 4200 in total). These cleaned texts were then 
concatenated to form one big input string of 148,214,028 characters. 

Entropies were estimated both from this string (which still contained upper 
and lower case letters, numbers, all kinds of brackets and interpunctation marks,
95 different characters in total), and from a version with reduced alphabet.
In the latter, we changed all letters to upper case; all brackets to either 
( or ); the symbols \$,\#,\&,*,\%, @ to one single symbol; colons, exclamation and 
question marks to points; quotation marks to apostrophes; and semicolons to commas.
This reduced alphabet had then 46 letters (including, of course, the blank
``$_\sqcup$").

The most frequent pair of letters in English is ``e$_\sqcup$". After replacing it 
by a new ``letter", the next pair to substitute is ``$_\sqcup$t", then ``$_\sqcup$a",
``$_\sqcup$th", etc. Very soon also longer strings are substituted, e.g. after 
92 steps appears the first two-word combination, ``of$_\sqcup$the$_\sqcup$".

As long as the number of new symbols is still small, it is easy to estimate the 
pair probabilities, and from this an upper bound $\hat{h}_i = h_i^{(2)}N_i/N_0$ 
on the entropy.  This becomes more and more difficult
as the alphabet size increases, as the sampling becomes insufficient even with 
our very long input file, and we can no longer approximate the $p_i(k,k')$ by the
observed relative frequencies. As long as the number of different subsequent pairs is 
much smaller than the sequence length (i.e., most pairs are observed many times), 
we can still get reliable estimates of $\hat{h}_i$ by using the leading correction 
term discussed in \cite{grass-fsc,footnote2}. But finally, when many pairs are seen only 
once in the entire text, we have to stop since any estimate of $h_i^{(2)}$ becomes 
unreliable.

We went up to 6000 substitutions. The longest substrings substituted by a single 
new symbol had length 13 in the original (95 letter) alphabet, and length 16 in the 
reduced (46 letter) one (the latter was ``would$_\sqcup$have$_\sqcup$been$_\sqcup$").
The entropies $\hat{h}$ per (original) character are plotted 
in Fig.3. We see that they are very similar for both alphabets.
We find $\hat{h}\approx 1.8$ bits/character after 6000 substitutions. This number 
is very close to the value obtained from most other methods (with the exception of 
\cite{teahan-cleary}, where $\approx 1.5$ bits/character were obtained), if one uses 
$10 - 100$ MB of input text \cite{bell,sg}. This is surprising in view of two facts.
First of all, the methods applied in \cite{bell,sg} are very different, and one 
might have thought a priori that they are able to use different structures of the 
language to achieve high compression rates. Apparently they do not. 

\begin{figure}
%Fig 3
\psfig{file=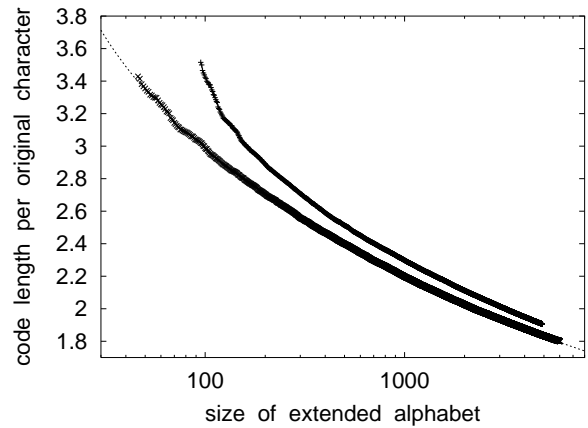,width=5.8cm,angle=270}
\caption{Entropy estimates $\hat{h}$ from pair probabilities plotted against 
   the size of the extended alphabet. Upper curve is for the initial 7 bit
   alphabet, including upper and lower case letters. The lower curve is for the 
   reduced (46 letter) initial alphabet. The smooth dotted line passing 
   through the lower data set is a fit with Eq.(\ref{fit}).}
\label{fig3.ps}
\end{figure}

Secondly, it is clear that $\hat{h}\approx 1.8$ bits/character is not a realistic
estimate of the true entropy of written English. Even though we can not, with our 
present text lengths and our computational resources, go to much larger alphabet sizes
(i.e. to more substitutions), it is clear from Fig.3 that both curves would continue
to decrease. Let us denote by $i$ the number of substitutions. Then empirical 
fits to both curves in Fig.3 are given by
\be
   \hat{h}_i = h + {c\over (i+i_0)^\alpha } \;.                \label{fit}
\ee
Such a fit to the 46 letter data, with $h=0.7, i_0=34, c=4.99,$ and $\alpha = 0.1745$, 
is also shown in Fig.3. One should of course not take it too serious in view of the 
very slow convergence with $i$ and the very long extrapolation, but it suggests that 
the true entropy of written English is $0.7\pm 0.2$ bits/character.

This estimate is somewhat lower than estimate of \cite{cov-king} and the 
extrapolations given in \cite{sg}. It is 
comparable with that of \cite{grass-ieee} and with Shannon's original estimate 
\cite{shannon2}. It seems definitely to exclude the possibility $h=0$ which was 
proposed in \cite{hilberg,ebel-posch}.

\section{Conclusions}

We have shown how a strategy of non-sequential replacements of pairs of characters
can yield efficient data compression and entropy estimates. A similar
strategy was first proposed by Jim\'enez-Monta\~no and others, but details and the 
actual coding done in the present paper are quite different from those proposed in 
\cite{jimenes,poeschl}. Indeed, this strategy was never used in \cite{jimenes,poeschl}
for actual codings, and it was also not used for realistic entropy estimates.

Compared to conventional sequential codes (such as Lempel-Ziv or arithmetic
codes \cite{cover}, just to mention two), the present method would be much 
slower. Instead of a single pass through the data as in sequential coding 
schemes, we had gone up to 6000 times through the data file, in order to 
achieve a high compression rate. We could do of course with much less passes,
if we would be content with compression rates comparable to those of commercial
packages such as ``zip" or ``compress". For written English these achieve typically
compression factors $\approx 2.6$, i.e. ca. 3 bits/character. As seen from Fig.1,
this can be achieved by NSRPS very easily with very few passes, but even then the 
overhead and the computational complexity of NSRPS is much too high to make it 
a practical alternative.

NSRPS can be seen as a greedy and extremely simple version of off-line textual 
substitution \cite{storer}. In combination with other sophisticated techniques, 
similar substitutions can give excellent results \cite{teahan-cleary}. But without
these techniques, it is in general believed that only much more sophisticated 
versions of off-line textual substitution are of any interest \cite{storer}.
Again this is presumably true as far as practical coding schemes are concerned.
But things seem to be different if one is interested in entropy estimation. Here the 
present method is much simpler (even though computationally more demanding) than 
the tree-based gambling algorithms \cite{sg,bell} that had given the best results
up to now. Without extrapolation, it gives the same (upper bound) estimates 
as these methods. But it seems that it allows a more reliable extrapolation to 
infinite text length and infinite substitution depth, and thus a more reliable
estimate of the true asymptotic entropy. 

From the mathematical point of view, we should however stress that we have only 
partial results. While we have proven that the Markov structure is a fixed point 
of the substitution, we have not proven that it is {\it attractive}. We thus 
cannot prove that the present strategy is indeed universally
optimal, although we believe that our numerical results strongly support this 
conjecture. A rigorous proof would of course be extremely welcome.

I thank Ralf Andrzejak, Hsiao-Ping Hsu, and Walter Nadler for carefully reading 
the manuscript and for useful discussions.


\begin{thebibliography}{99}
\bibitem{footnote2} We use Eq.(13) of \cite{grass-fsc}, but with a misprint 
   corrected: The denominator of the last term should be $(n_i+1)n_i$ instead 
   of $n_i+1$.
\end{thebibliography}
\end{document}